\PassOptionsToPackage{unicode}{hyperref}
\PassOptionsToPackage{hyphens}{url}
\PassOptionsToPackage{numbers,sort&compress}{natbib}
\documentclass[
]{article}
\usepackage{comment}
\usepackage{physics}
\usepackage{setspace}
\usepackage[export]{adjustbox}
\usepackage{fullpage}
\usepackage{natbib}
\usepackage{caption}
\usepackage{subcaption}
\usepackage{amsmath,amssymb}
\usepackage{lmodern}
\usepackage{graphicx}
\usepackage{iftex}
\usepackage{comment}
\ifPDFTeX
  \usepackage[T1]{fontenc}
  \usepackage[utf8]{inputenc}
  \usepackage{textcomp} 
\else 
  \usepackage{unicode-math}
  \defaultfontfeatures{Scale=MatchLowercase}
  \defaultfontfeatures[\rmfamily]{Ligatures=TeX,Scale=1}
\fi
\IfFileExists{upquote.sty}{\usepackage{upquote}}{}
\IfFileExists{microtype.sty}{
  \usepackage[]{microtype}
  \UseMicrotypeSet[protrusion]{basicmath} 
}{}
\makeatletter
\@ifundefined{KOMAClassName}{
  \IfFileExists{parskip.sty}{%
    \usepackage{parskip}
  }{
    \setlength{\parindent}{0pt}
    \setlength{\parskip}{6pt plus 2pt minus 1pt}}
}{
  \KOMAoptions{parskip=half}}
\makeatother
\usepackage{xcolor}
\IfFileExists{xurl.sty}{\usepackage{xurl}}{} 
\IfFileExists{bookmark.sty}{\usepackage{bookmark}}{\usepackage{hyperref}}
\hypersetup{
  pdftitle={Untitled},
  pdfauthor={aurthor},
  hidelinks,
  pdfcreator={LaTeX via pandoc}}
\ifLuaTeX
  \usepackage{selnolig}  
\fi

\title{Noise as a probe of Majorana fermion surface states}
\author{Saleem Al Dajani$^{\dag,1, *}$ \& Henning Soller$^{\dag, 2, *}$ \\ \small \textit{$^{\dag}$These authors\footnote{Corresponding authors: sdajani@mit.edu, henning\_soller@mckinsey.com} contributed equally to this work} \\ \scriptsize \textit{$^1$ Massachusetts Institute of Technology, KAUST, UCB-NE?} \\ \scriptsize  \textit{$^2$ McKinsey\&Company, University of Heidelberg} }
\date{January 16, 2022}

\begin{document}

\maketitle

\abstract{
The emergence of Majorana fermion surface states has been studied in several environments. In this paper we discuss the current-voltage characteristics of a Majorana fermion state as a function of the voltage, temperature and magnetic field and compare the results to experiment. Based on the agreement and identified parameters we calculate the noise as a possible ultimate proof of the emergence of Majorana fermions.}

\section{Introduction}

Shortly after the relativistic wave equation was derived and the antiparticle was predicted in 1928 by Paul Dirac, \cite{dirac1936relativistic} Ettore Majorana theorized in 1937 that it also describes ‘Majorana’ fermions that are their own antiparticles. \cite{majorana1937theory} Recently, Majorana fermions have emerged as building blocks for noise-free quantum computers due to their topological nature \cite{cheng2012topological, flensberg2021engineered, beenakker2013search}. Majorana fermion surface states have been studied in several environments, including a ferromagnetic material in contact with a superconductor exhibiting high spin-orbit coupling, such as in 'Majorana islands' of europium sulfide in contact with superconducting gold nanowires on vanadium \cite{manna2020signature, wei2020demonstration}.

In high-energy particle physics, the Majorana nature of neutrinos are also under investigation in $^{76}$Ge via the detection of neutrinoless double-beta ($\beta$) decay ($0\nu \beta \beta$) using high purity Germanium (HPGe) detectors to verify whether the total lepton number is violated, which would also prove the existence of the Majorana nature of subatomic particles — in this case, of Majorana neutrinos. This would validate the neutrino mass scale by a seesaw mechanism and bound the neutrino absolute mass. \cite{elliott2015colloquium, aalseth2018search, klapdor2001latest, aalseth2000recent, klapdor2004search, abgrall2018processing, xu2015majorana, guiseppe2018new} With low enough electronic noise, these experiments may also enable a dark matter search for weakly interacting massive particles (WIMPs) and axions with masses below 10 GeV to resolve issues in quantum chromodynamics (QCD) \cite{giovanetti2015dark}. Measuring the noise in these decay processes may also yield more conclusive results via a full counting statistics approach. 

The typical sign for the Majorana fermions in superconductor heterostructures has been the emergence of a zero-bias peak in current-voltage characteristics - which could, however, also be the result of various other phenomena such as the Kondo effect \cite{soller2014charge}. Generating Majorana fermion nodes in superconductors enables the study and analysis of these topologically protected states with tunable parameters in a controlled environment \cite{sdajani2022}. This paves the way for the usage of Majorana fermion nodes as building blocks for quantum computers, as their emergence as a topological state should provide for significant protection of these states \cite{kitaev2001unpaired}.

In this paper, we would like to analyze the specific situation for gold
surface states and possible descriptions of the underlying system.
Additionally, we would like to compare them in section II to recent
experimental results and then reflect in section II on the possibility of
additional measurements to prove the nature of the Majorana fermion node
before concluding in section IV.

\section{Theoretical description and background}

Majorana fermions can be realized by analyzing an electron-hole pair in a double quantum well. \cite{refael2015} In this case, the presence of protected topological states emerge as a result of spin-orbit coupling, exchange, and superconductivity. \cite{oreg2010helical} These bound states can be predicted by employing Hamiltonian formalism using full counting statistics \cite{soller2014charge}. The presence of Majorana quasiparticles may be detected with a myriad of experimental techniques, with some recent studies that have shown preliminary, yet unconclusive evidence \cite{manna2020signature, wei2020demonstration}.

Analytically modeling Majoranas requires a derivation of a Hamiltonian to model topologically protected states in a system with the right properties \cite{kitaev2001unpaired}. Measuring conductance in these systems yields a zero bias peak \cite{wei2020demonstration,manna2020signature}, however, this is not unique to Majorana fermions.

Measuring noise should yield a second oberservable that should clarify the charge transfer mechanism and therefore a conclusive measure of their existence. \cite{sdajani2022} Other phenomena that yield a zero bias peak in conductance such as the Kondo effect should proceed via single electrons and therefore show a different noise behavior. Numerous methods are available to measure the noise also in topological systems. \cite{beenakker2013search, jack2021detecting}

In addition to the Hamiltonian formalism numerical simulations provide a way to study the conductance in these systems. The numerical analysis proceeds as the Hamiltonian approach via a Green's function self-energy. \cite{manna2020signature} The Kubo formula may also be applied to obtain the noise from the current-current correlation function. \cite{bruus2004many} Further analysis of the noise in the system via the Keldysh formalism also yields the noise signature of Majoranas in platforms for topological qubits. \cite{liu2015probing, burtzlaff2015shot}

Simplified models may also be used to simulate the behavior of Majorana fermions in topologically coupled chains with tight-binding theory, where the time-independent Schr\"{o}dinger equation is discretized and parametrized by hopping parameters. \cite{groth2014kwant} Further complexity may be added to study these phenomena via tight-binding, such as Majoranas bound to vortices in a superconducter-topological insulator 3D model. \cite{papaj2021creating} 

The understanding of the fundamental transport properties of Majoranas are important for the development of topological qubits that rely on Majorana fermions for transfer of information from one node to another. \cite{steiner2020readout} This is done by storing quantized information in topologically-protected Majorana states. \cite{cheng2012topological} With the recent developments in artificial intelligence and machine learning using high-performance computational infrastructure, qubits can be initialized, emulated, and controlled in an unprecedented fashion. \cite{coopmans2021protocol}

The theoretical description employed may be extended to exciton condensates formed as topological excitons in systems with strong spin-orbit coupling, such as perovskites. \cite{davis1976proposed, little1981criteria, miller1976experimental, bardeen1978excitonic} This paves the way for the use of exciton condensates towards the design of excitonic superconductors and insulators in long life-time, charge-free functional devices, as well as excitonic superconductors in magnetic tunnel junctions, lasers, solar cells, light-emitting diodes, batteries, instruments, detectors and other devices with efficiencies that, under the right operational conditions and with suitable fabrication protocols (i.e. overcoming issues like contact resistivity), approach unity. 

We will describe the aforementioned based on EuS as a superconducting system in contact with leads following the Hamiltonian approach \cite{cuevas1996hamiltonian}. Specifically, we will follow the prior considerations for the description of a Majorana fermion in a superconductor
environment involving a ferromagnet \cite{soller2014charge}. The effective low-energy behavior
of the system is that of a $p$-wave superconductor and we can effectively
treat the system described on the superconducting gold surface state
using the Hamiltonian

\begin{eqnarray}
H_{eff} = \sum_{k}\xi_{k} \psi_{k\uparrow}^{+} \psi_{k\uparrow} + \sum_{k}(\Delta_{p} \psi_{k\uparrow}^{+} \psi_{-k,\uparrow}^{+} +\Delta_{p}^{*} \psi_{-k, \downarrow} \psi_{k,\downarrow})
\label{eq1}
\end{eqnarray}

where \(\Delta_{p}\) refers to the effective temperature dependant
$p$-wave gap. we will assume the validity of the high temperature limit of
the gap equation.

\[\Delta_{p}(T) = \Delta_{p}(1- T/T_e)^{1/2}\]
\[\Delta_{p}(B) = \Delta_{p}(1- B/B_e)^{1/2}\]

The Hamiltonian in Eq. (\ref{eq1}) is very similar to an $s$-wave
superconductor Hamiltonian so that we can easily take over previous
results from the studies of $s$-wave superconductors \cite{muzykantskii1994quantum}. The tip of the
scanning tunneling microscope can be described as a normal metal with a
flat bend density of states \(\rho_{OT}\) using field operators
\(\psi_{T,k,\delta}\)

\[H_{STM} = \sum_{k,\sigma} \epsilon_{k}\psi_{T,k,\sigma}^+ \psi_{T,k,\sigma} \]
The STM is held at chemical potential \(\mu_T\) whereas we keep the
gold surface state at \(\mu = 0\) in accordance with previous studies of
SC point control \cite{buitelaar2003multiple}. The description of the tunneling between the surface
state and the STM tip is given by the usual tunneling Hamiltonian.
We need to include the possibility of an additional phase shift \(\phi\)
during the Andreev reflection\cite{soller2014charge}. The phase shift needs to be explicitly
taken into account given that Andreev reflection is a coherent process
involving a hole and an electron so that it may appear in physical
results.

\begin{eqnarray}
H_T =\gamma_T[e^{i \tilde{\phi}/2}\psi_{\uparrow}^{+}(x=0)\psi_{T\uparrow}(x=0)+h.c.]
\label{eq2}
\end{eqnarray}

This way the additional phase shift \(\phi\) accounts for the
topological phase of the system. For simplicity we choose
\(\tilde{\phi}=_{-}^{+}\phi\) per electron/ hole.

For this system in question we do not really want to describe the
current but the full behavior of the system which is described by the
full counting statistics \cite{belzig2001full}. This is the Fourier transform of the probability
distribution fuction $P(Q)$ of transferring $Q$ units of charge during a
given (long) measurement time $\tau$. Physical observables can then be
calculated as average with respect to this distribution function.

However, instead of directly calculating \(P(Q)\) it is often more
convenient to calculate the cumulant generating function (CGF)
$\chi(\lambda) \quad = ln \sum_Q e^{i\lambda Q} P(Q)$. This allows to
calculate the $\chi(\lambda)$ via Keldysh Green's function\cite{gogolin2006towards} following

\begin{eqnarray}
\chi(\lambda)=  \langle T_C e^{-i \int_C dt T^\lambda (t)}\rangle_0,
\end{eqnarray}
where \(T^{\lambda(t)}\) refers
to the tunnel Hamiltonianin Eq. \ref{eq2} with the substitution
\(\psi_\uparrow(x=0) = \psi_\uparrow e^{-i\lambda/2}\).

We can now follow the Hamiltonian approach to calculate the CGF as a sum
of two contributions referring to the behavior above and below the $p$-wave
gap. This should also allow to identify the individual charge transfer mechanisms.

Introducing the phase difference \(\phi=\pi\) in Eq. \ref{eq2} leads to the
typical behavior of the Andreev bound state merging into one bound state
at $\epsilon_{mBS}= \Delta_p \cos{\pi/2}=0$\cite{beenakker2013search}. The full expression for
the CGF reads

\begin{eqnarray}
\chi(\lambda) &=& \chi_e(\lambda) + \chi_A(\lambda)) \nonumber \\
&=& \tau \gamma_T \int \frac{d\omega}{2\pi} ln \{1 + T_A(\omega)[(e^{2i\lambda}-1) n_T(1-n_{T+})(e^{2i\lambda}- 1)n_{T+}(1 -n_T)]\} \nonumber \\
&& + \tau \gamma_T \int \frac{d\omega}{2\pi} ln \{1+T_e(\omega)[(e^{i \lambda} - 1) n_T (1 -n_F)+ (e^{-i \lambda} - 1) n_F (1-n_T)]\}\nonumber \\
&& T_e(\omega)= \frac{4 \Gamma |\omega|}{\sqrt {\omega^2-\Delta_p^2}}\qquad, \Gamma= (1+P) \pi^2 \rho_{0T} \rho_0 \gamma_T^2 \nonumber\\
&& T_A(\omega)= \frac{T^2}{1+R^2-2R\cos{2 \arccos{\omega/\Delta_p +\pi \theta(B/B_T -1)}}} \nonumber\\
&& R=1-T, \; T=4\Gamma, \; n_{T+}=1-n_T(-\omega) \label{eq3}
\end{eqnarray}
This model allows to identify whether the noise and current
characteristics actually replicate the behavior of the system in the
experiment. The transmission coefficients correspond to the behavior of the conductance at zero temperature so that for zero-bias conductance $G(V=0) = G_0 T_A(\omega)/2$.

\section{Temperature Dependence}
\begin{figure}[!ht]
	\centering
	\includegraphics[width=0.9\textwidth]{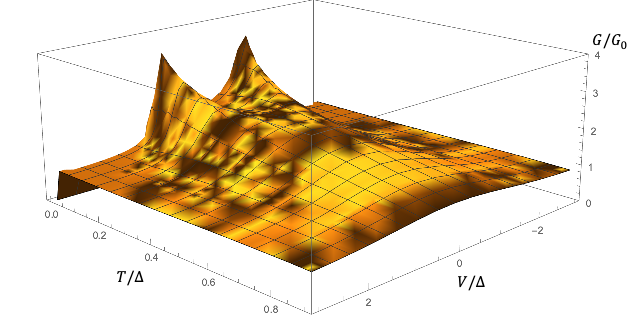}
	\caption{Overview of the conductance as a function of temperature $T$ and voltage $V$ below the critical magnetic field. We see the conductance peaks disappearing at higher voltages and temperatures.}
	\label{fig1}
\end{figure}

Deriving the current as the first derivative
with respect to $\lambda$ of Eq. (\ref{eq3}) and calculating the conductance
$G= dI/dV$ we obtain in Fig. \ref{fig1} the same behavior as in
the experiment below the magnetic field $B_T$. When going below $B_T$ we see the usual two-peak
structure of the Andreev bound state that disappears upon rising the temperature above \(T_c\). This behavior is to be expected as the superconductivity is vanishing for large enough temperatures as we have also observed in the experiment.

\section{Magnetic Field Dependence}
\begin{figure}[!ht]
	\centering
	\includegraphics[width=0.9\textwidth]{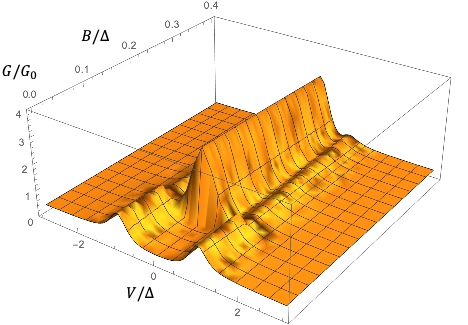}
	\caption{Overview of the conductance as a function of magentic field $B$ and voltage $V$. We see the emergence of the Majorana conductance peak when going beyond $B>B_T=0.1\Delta$.}
	\label{fig2}
\end{figure}

Deriving the current as a function of the
magnetic field in Fig. \ref{fig2} we see that below $B_T<B_C$ the behavior has the two
peaks aligned to the Andreev levels. Upon raising $B>B_T$ the Majorana
peak emerges with sidebands arising from the normal conducting behavior
of the superconductor density of states at $V>\Delta_p$. The Andreev
states are slowly moving in as the superconductors $\Delta_p$ depends
also on the magnetic field $B$. This has also been observed in the experiement as the Majorana leads to the peak arising to almost perfect conductance and the Andreev levels vanish when the Andreev bound states merge beyond the critical magnetic field.

Both of these observations are in
perfect alignment with the experiment. Specifically we see the right
behavior of a sudden and steep jump of the conductor at \(V=0\) close to
the theoretical value of \(G = 2G_0\) (perfect conductivity).

\section{Numerical Simulations}
For a numerical simulation of the system, a surface-state (SS) Green's function (GF) approach is applied to describe surface-bulk mixing. The surface bulk mixing is assumed to be due to impurity scattering at a Matsubara frequency, $\omega_n$, and strength, $\Gamma$. The effective action is then averaged over disorder configurations within the Born approximation to derive the self-energy, $\Sigma(i\omega_n)$. This self-energy term is then incorporated into the SSGF to obtain the gold surface-state effective Hamiltonian when in contact with EuS islands on vanadium under a magnetic field parallel to the surface:  
\begin{equation}
    H_{eff}=Z(\frac{k^2}{2m}-\mu)\tau_z + Z \alpha_R(\mathbf{k}\times\mathbf{\sigma})\cdot\hat{z}\tau_z + (1-Z)\Delta_B \tau_x + \mathbf{V}\cdot\mathbf{\sigma}\tau_0
\end{equation}
where $\tau_i$ are the Pauli matrices that describe the particle-antiparticle pairs, $\sigma$ is the spin, $\Delta_B$ is the superconducting gap, $k\times \sigma$ is the exchange term, and $V\cdot \sigma\tau_0$ is the spin-orbit coupling (SOC) term. Taken altogether, a Majorana bound state (MBS) at the edges of the gold nanowire are described by this model and can be probed by a scanning tunneling microscope (STM) in contact with the gold surface. 

\begin{figure}[!ht]
	\centering
	\begin{subfigure}{0.45\textwidth}
         \centering
\includegraphics[width=\linewidth, left]{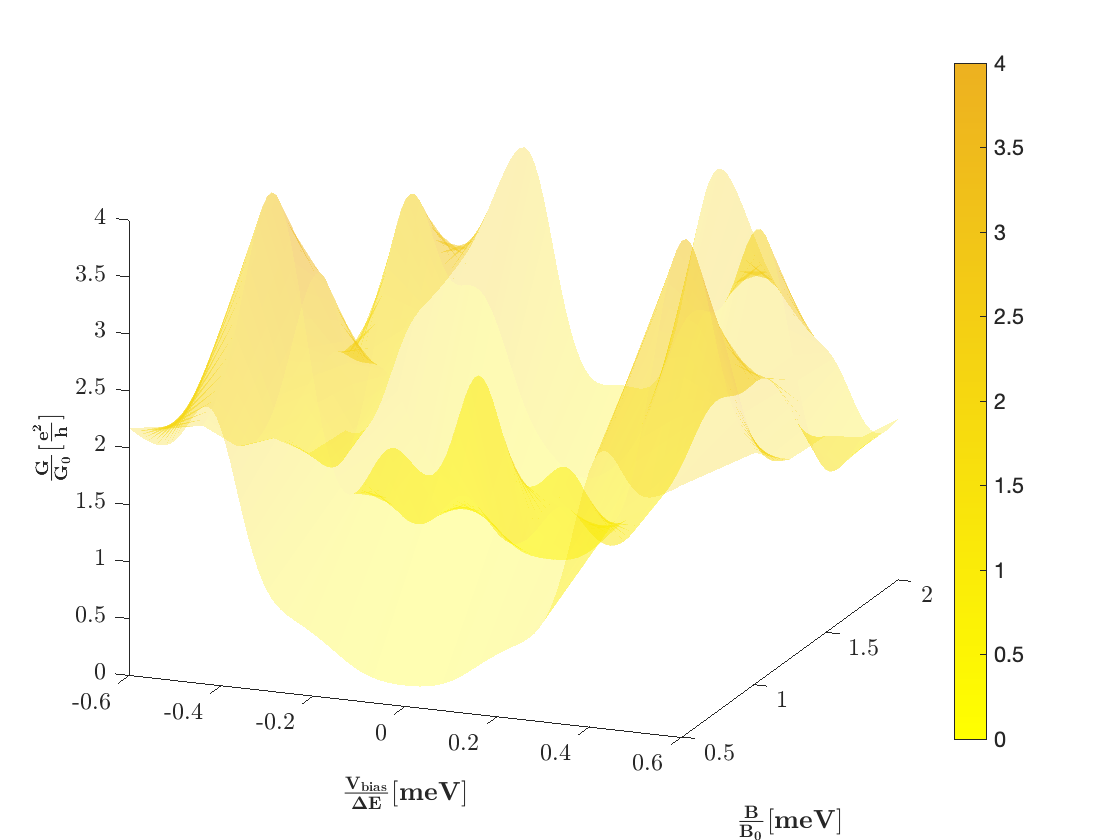}
         \caption{3D-view}
         \label{fig3a}
     \end{subfigure}
     \hfill
     \begin{subfigure}{0.45\textwidth}
         \centering
\includegraphics[width=\linewidth, right]{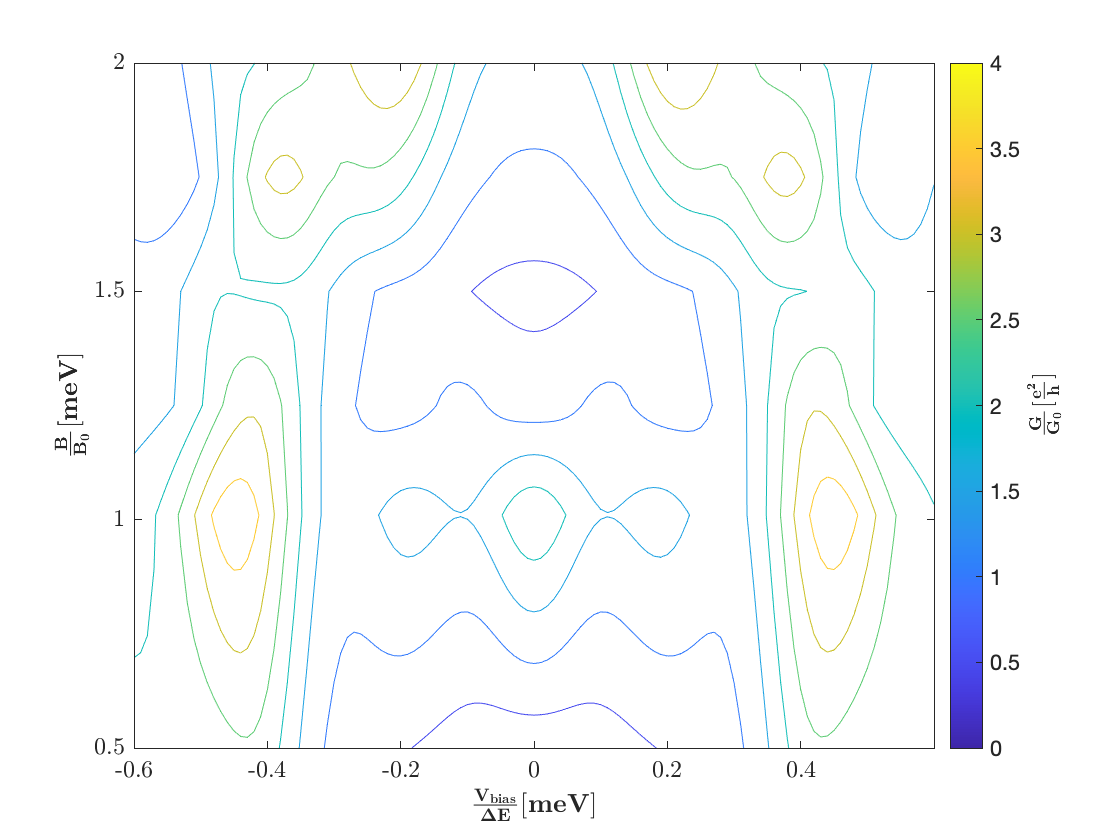}         \caption{Contour-view}
         \label{fig3b}
     \end{subfigure}
	\caption{Tuning conductance of zero-bias surface MBSs via
magnetic field by numerically simulating surface-bulk mixing. (a) 3D-view illustrates zero bias Majorana zero mode peak in agreement with analytical model. (b) Contour-view highlight the critical magnetic field at which the zero bias peak is observed in agreement with experiments reported in \cite{wei2020demonstration, manna2020signature}. At the zero bias peak, the calculation yields a value of two corresponding to the pair of Majorana edge states (MES) as a result of the formation of an MBS under these conditions. }
	\label{fig3}
\end{figure}

The conductance calculated from this effective Hamiltonian as a function bias voltage and its temperature and magnetic field dependence is in agreement with the analytical results. A representative figure of the magnetic field dependence is shown in Fig. \ref{fig3}

To elucidate the origin of the zero bias peak in conductance and the mechanism responsible for its emergence, a tight-binding approach is employed \cite{groth2014kwant}. A multi-dimensional version of the time-independent Schr\"{o}dinger equation (TISE) is put forth:
\begin{equation}
    H_{TB}=\frac{-\hbar^2}{2m}(\nabla^2) + V
\end{equation}

For the 2-D case, this equation parameterized with a single hopping parameter, $t$ and then discretized to the lattice structure, by a lattice constant $a$, specified by the system in question: 

\begin{eqnarray}
H_{TB}^{2D}&=&\frac{-\hbar^2}{2m}(\partial_x^2 + \partial_y^2) + V(y) \\
    t &=& \frac{\hbar^2}{2ma^2} \\
    \ket{i,j} &=& \ket{ai, aj} = \ket{x, y} \\
    \partial_y &=& \frac{1}{a^2}\sum\limits_{i,j}(\ket{i+1,j}\bra{i,j} + \ket{i,j}\bra{i+1,j} -2\ket{i,j}\bra{i,j})
\end{eqnarray}

The algorithm is then implemented under the limit as $a \longrightarrow 0$ with a discretized TISE
including the on-site Hamiltonian with electron hopping in the x-direction and y-direction, such that the Hamiltonian describing the system is given by:

\begin{equation}
    H=\sum\limits_{i,j}[(V(ai,aj)+4t)\ket{i,j}\bra{i,j} - t(\ket{i+1,j}\bra{i,j} + \ket{i,j}\bra{i+1,j} + \ket{i,j+1}\bra{i,j} + \ket{i,j}\bra{i,j+1}]
\end{equation}

\begin{figure}[!ht]
	\centering
	\begin{subfigure}{0.45\textwidth}
         \centering
\includegraphics[width=\linewidth, left]{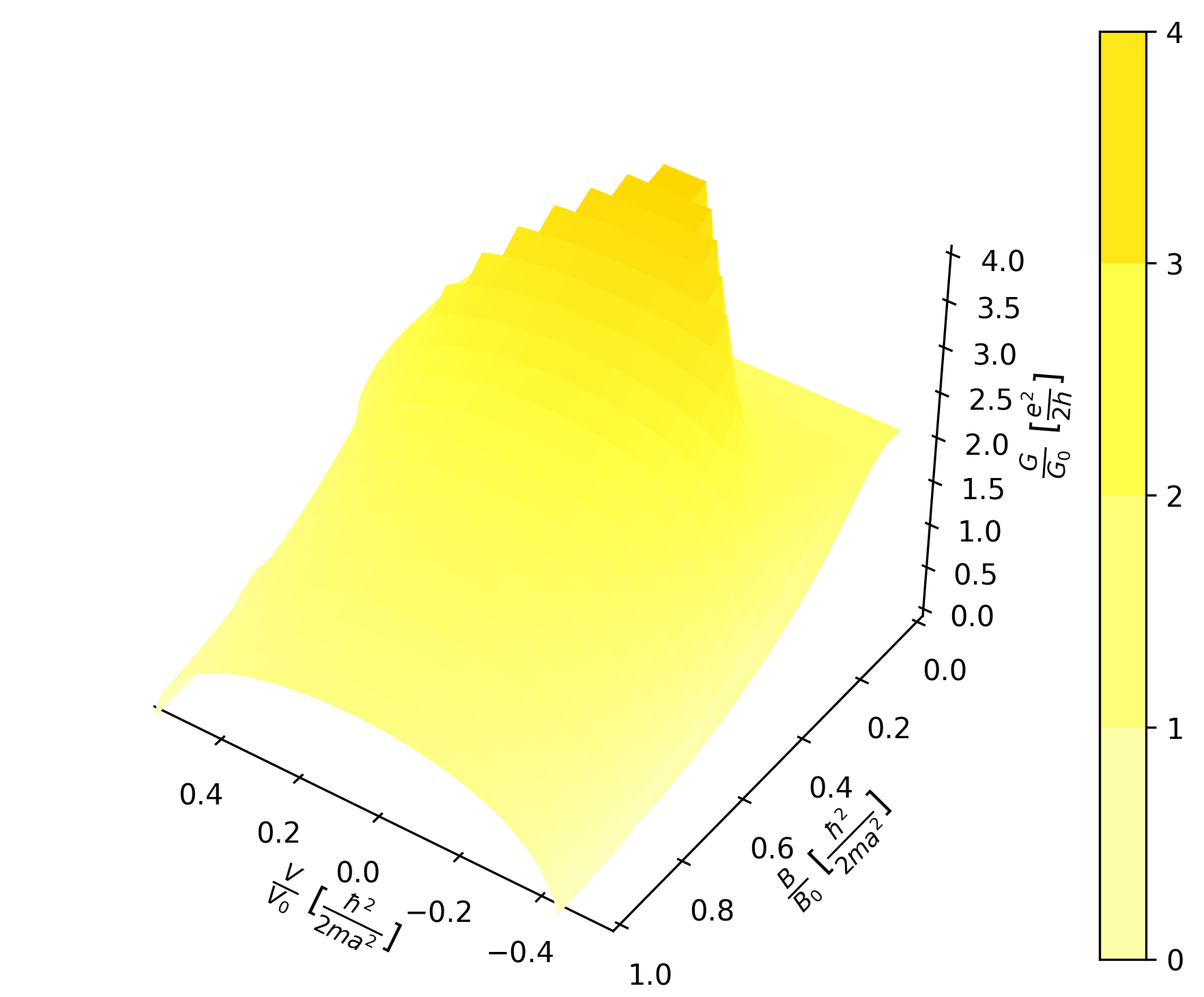}
         \caption{3D-view}
         \label{fig4a}
     \end{subfigure}
     \hfill
     \begin{subfigure}{0.45\textwidth}
         \centering
\includegraphics[width=\linewidth, right]{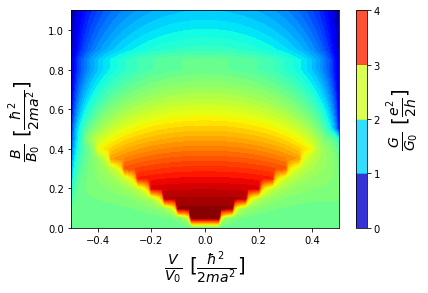}         \caption{Contour-view}
         \label{fig4b}
     \end{subfigure}
	\caption{Tuning conductance of zero-bias surface MBSs via
magnetic field by modeling a Kitaev chain with tight binding. (a) 3D-view illustrates zero bias Majorana zero mode peak in agreement with the analytical model and numerical simulation. (b) Contour-view highlight the critical magnetic field at which the zero bias peak is observed in agreement with experiments reported in \cite{wei2020demonstration, manna2020signature}. The value of four quantum units of conductance for the MBS indicates contributions from the pair of Majorana edge states (MES) as well as the superconducting Cooper pair states to the overall signal at the nanowire leads.}
	\label{fig4}
\end{figure}

Hard-wall 'particle in a box' confinement is achieved by limiting hopping
beyond a certain spatial region and in empty lattice sites. With this approach, the conductance in a Kitaev chain is calculated to confirm the emergence of a zero bias peak, and shown in Fig. \ref{fig4}. The fact that the MBS behavior in this system may be modelled by tight-binding implies that the chemistry of the system in question, i.e. the EuS/Au/V junction, is negligible and the physics dominate the emergence of this phenomenon.

\section{Noise Predictions}

\begin{figure}[!ht]
	\centering
	\includegraphics[width=0.9\textwidth]{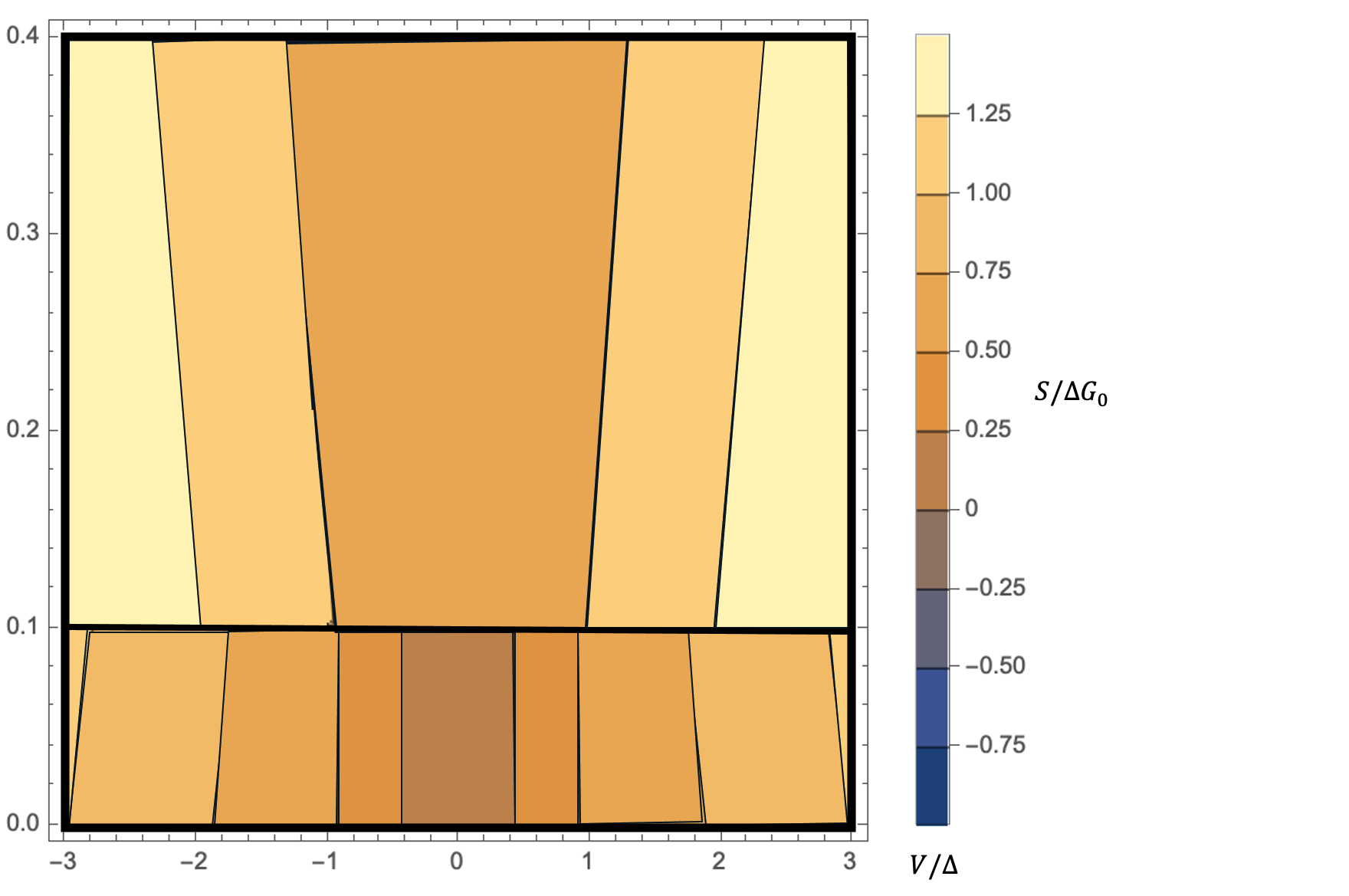}
	\caption{Overview of the noise as a function of voltage and magentic field. We see a steep jump at $B=B_T$ for the noise but it is not as steep as the jump in conductance in Fig. \ref{fig2}.}
	\label{fig5}
\end{figure}

This study gives us confidence that we
should be able to make a proper prediction for the noise as the second
derivative of $\chi(\lambda)$ with respect to \(\lambda\). We have
plotted \(S\) as a function of voltage and magnetic field, see Fig. \ref{fig5}.

We observe the noise following the behavior of the current meaning that
the additional noise contribution decreases massively at \(V=\Delta\) below \(B=B_T\) as this is the
position of the Andreev bound states and where the system becomes normal
conducting. This is the perfect reflection of \(S=2eI\) for Andreev
reflection. Above \(B_T\) we see that the noise is rising as the system
becomes more transparent and conducts better but the rise does not
reflect the massive change in conductivity seen before. This is due to
the fact that due to the perfect conductivity of the Majorana the usual
Fano factor of 2 for Andreev reflection below the gap does not apply.
This means that the noise is indeed a clear differentiator also for the
parameters chosen in the experiment to differentiate a Majorana fermion
from other possible effects leading to a steep rise in conductance below
the superconductor gap.

\section{Conclusion}

We have analyzed the behavior of a superconducting system
including a Majorana fermion theoretically in comparison to
experimental data.

The study has given us confidence that the Hamiltonian formalism allows
for a proper description of the system in question. It has also revealed
the noise features of the system showing that the typical Fano factor of
2 should be drastically reduced for Andreev reflection off a Majorana
fermion also for experimentally observable parameters.

This behavior can be measured experimentally using techniques previously employed for the study of normal-superconductor point contacts and there noise behavior.

\bibliographystyle{unsrt}
\begingroup
\setstretch{0.8}
{\scriptsize  
\bibliography{trial.bib}}
\endgroup

\end{document}